\documentclass[a4paper,11pt]{article}
\usepackage{jheppub} 
\usepackage{lineno}
\usepackage{framed}
\usepackage{color}
\usepackage{makecell}

\title{\boldmath Unified Origin of Curvature Perturbation and Baryon Asymmetry of the Universe}

\author[a]{Anish Ghoshal,}
\author[b]{Abhishek Naskar,}
\author[c]{and Nobuchika Okada}
\affiliation[a]{Faculty of Physics, Institute of Theoretical Physics,
University of Warsaw, ul. Pasteura 5, 02-093 Warsaw, Poland}
\affiliation[b]{School of Physical Science and Technology, ShanghaiTech University, 393 Middle Huaxia Road, Pudong, Shanghai 201210, China}
\affiliation[c]{Department of Physics and Astronomy, University of Alabama,
Tuscaloosa, Alabama 35487, U.S.A.}

\emailAdd{anish.ghoshal@fuw.edu.pl, naskara@shanghaitech.edu.cn, okadan@ua.edu}

\abstract{
We propose a unified framework that describes both the curvaton mechanism for generating primordial density fluctuations and the Affleck-Dine (AD) mechanism for baryogenesis. By introducing a complex scalar field (AD field) carrying a baryon/lepton number and its potential consisting of quadratic and quartic terms with a small baryon/lepton-number-violating mass term, we investigate the evolution of the scalar field during the radiation-dominated era following inflation.
We set the initial conditions such that the quartic term dominates the scalar potential, 
and the angular component of the AD field is non-zero. 
We focus on a scenario where the AD field sufficiently dominates the energy density of the universe before its decay.
We show that the radial component of the AD field can be identified with the curvaton
to solely produce the Planck normalized scalar power spectrum while the evolution of the angular component 
is crucial for generating the observed baryon asymmetry of the universe. 
Additionally, we find that the amplitude of scalar bispectrum $f_{NL}$ is negative, 
which is consistent with the current Planck data and testable in future observations such as CMB-S4, LiteBIRD, LSS, and 21-cm experiments. 
In our estimation of the scalar power spectrum and bispectrum, we develop a novel analytical scheme for computing scalar fluctuations 
based on the $\delta N$ formalism, which allows us to deal with the evolution of curvaton with polynomial potential more accurately in comparison to the existing analytical methods.
}

\makeatletter
\gdef\@fpheader{}
\makeatother

\begin{document}
\maketitle
\flushbottom

\section{Introduction}\label{sec:intro}
The observed temperature fluctuation in Cosmic Microwave Background (CMB) radiation is often attributed to the quantum fluctuation of the inflaton field which is responsible for cosmic inflation. This observed fluctuation can also be explained with other mechanisms and one of the most explored mechanisms is the curvaton scenario \cite{Linde:1996gt,Lyth:2001nq,Lyth:2002my,Lyth:2005qk,Lyth:2006nx,Enqvist:2009ww}. In this scenario, the curvaton field does not influence the inflation dynamics and stays as a spectator field but produces isocurvature perturbation through its quantum fluctuations. After inflation ends and reheating is completed, initially the curvaton energy density remains sub-dominant with respect to the radiation. The radiation energy density redshifts away faster than the curvaton energy density and the curvaton field converts the isocurvature perturbation into observable curvature perturbation at the time of its decay. The contribution of the curvaton energy density to the total energy density of the universe can be subdominant or dominant  at the end of its decay depending on its decay width, leading to different observational consequences.

Recently the curvaton mechanism has been explored in various interesting cosmological scenarios such as scalar perturbations with large non-Gaussianity generally associated with the large decay width of the curvaton \cite{Langlois:2008vk,Byrnes:2011gh,Kawasaki:2012gg,Kawasaki:2011pd}, primordial gravitational waves \cite{Kawasaki:2021ycf,Kawasaki:2013xsa,Franciolini:2023pbf,Inomata:2023drn} and primordial black holes \cite{Pi:2021dft,Ferrante:2023bgz,Chen:2023lou,Chen:2024pge,Inomata:2020xad,Chen:2019zza}. In this article we explore the curvaton mechanism to explain another mystery of cosmology: the origin of matter-anti-matter asymmetry. For our purpose we consider the Affleck-Dine (AD) mechanism \cite{Dine:1995uk,Affleck:1984fy} where an evolution of a scalar field carrying baryon/lepton number in the early universe generates  baryon/lepton asymmetry. A unified picture of the inflation and AD mechanism has been proposed  \cite{Cline:2019fxx,Cline:2020mdt,Charng:2008ke,Hertzberg:2013mba,Takeda:2014eoa,Lin:2020lmr,Kawasaki:2020xyf,Lloyd-Stubbs:2020sed} where the AD field also serves as the inflaton field. In this article, we explore a unification of the AD field with the curvaton that simultaneously produces the observed scalar perturbations and the baryon asymmetry of the universe.

The form of the curvaton potential energy plays a crucial role in the computation of the primordial observables. The $\delta N$ formalism is a powerful tool to compute the observables \cite{Sasaki:1995aw}. For a curvaton scenario with a quadratic potential (known as the vanilla curvaton scenario) the observables can be estimated analytically \cite{Kawasaki:2011pd,Ellis:2013iea,Matsuda:2007av}. There are several developments in the computations of observables from curvaton with more complicated potentials, e.g. polynomial potential \cite{Byrnes:2011gh,Huang:2008zj}, axion type potential \cite{Kawasaki:2011pd,Kobayashi:2020xhm} and multiple curvaton scenario \cite{Huang:2008rj}. In this article we take a different method for analytic estimation (for an alternate method see \cite{Kawasaki:2011pd}). For our analysis we consider a curvaton potential with two terms: $\frac{1}{2} m^2 \phi^2 + \frac{1}{4} \lambda \phi^4$, where $m$ is the mass of the curvaton, and $\lambda$ is the quartic coupling constant \footnote{Our method can also be applied to potentials like $V(\phi) = \frac{1}{2} m^2 \phi^2 + \sum_{n>2} \frac{1}{n} \lambda_n \phi^n$, which is explored in \cite{CurvatonMethod}.}. We split our system into three separate regions: 1.~curvaton field oscillation with its potential dominated by the quartic term, while the radiation energy density dominates; 2.~curvaton field transits to the quadratic region at a transition field value $\phi_T$ computed by the condition $\frac{1}{2} m^2 \phi_T^2 = \frac{1}{4} \lambda \phi_T^4$. In this region radiation energy density still dominates over curvaton energy density; 3.~curvaton field oscillation with its potential dominated by the quadratic term until its decay. Using the behavior of curvaton energy density in quartic and quadratic region \cite{Cembranos:2015oya} we compute the post-inflationary Hubble parameter in these three regions using appropriate boundary conditions, and then compute the post inflationary number of e-folds with which the relevant observables can be computed. Note that the above discussion about the system is valid if the initial curvaton field value $\phi_I$ is greater than the transition field value $\phi_T$. If $\phi_I < \phi_T$, our analysis approximately reduces to the vanilla curvaton scenario. To check the consistency of our approach we have computed all the known results for the curvaton scenario with a quadratic potential in Sec.~\ref{sec:SHO}. 

The curvaton potential we consider in this article is essentially the same as the one considered in \cite{Lloyd-Stubbs:2020sed}, where the AD field is identified as the inflaton. In \cite{Lloyd-Stubbs:2020sed} a non-minimal coupling with gravity is introduced for the AD field to drive inflation. However, the effect of the non-minimal coupling becomes negligible for the AD field evolution after the end of inflation. In this article we consider the AD field to be a curvaton, without any non-minimal coupling with gravity, that produces both the scalar perturbations and baryon asymmetry. Our transition scheme for the curvaton from quartic to quadratic region of the potential is similar to the threshold approximation discussed in \cite{Lloyd-Stubbs:2020sed,Babichev:2018sia}. As the AD field is a complex scalar field, we can decompose it into a radial field and an angular field. The radial field behaves as the curvaton with the potential $\frac{1}{2} m^2 \phi^2 + \frac{1}{4} \lambda \phi^4$ while the angular field is almost frozen with negligible energy density and does not contribute to the generation of scalar perturbations. Although the angular field negligibly affects the dynamics of the radial field, its evolution is crucial to produce sufficient amount of baryon asymmetry of the universe.

This article is organised as follows: in Sec.~\ref{sec:deltaN} we briefly discuss the $\delta N$ formalism; in Sec.~\ref{sec:Method} we develop our method for quadratic potential and quadratic + quartic potential curvaton scenarios; in Sec.~\ref{sec:smallR} we analyze the curvaton scenario with a large decay width and explore the possibility of producing observable non-Gaussianity; in Sec.~\ref{sec:ADbaryo} we analyze the AD curvaton mechanism and discuss the possibility of observable baryogenesis. 
The last section is devoted to conclusions and discussions. 

\section{$\delta N$ Formalism and Scalar Perturbation}\label{sec:deltaN}

We employ the $\delta N$ formalism \cite{Sasaki:1995aw} to compute the scalar power spectrum and bispectrum.  The quantity $\delta N$ is defined as the  perturbation in the number of e-folds between a spatially flat slice and a uniform density slice, and the curvature perturbation depends linearly on $\delta N$. If there are $P$ different fields, we can write down the curvature perturbation in terms of the fields' fluctuations as \cite{Sasaki:1995aw}
\begin{align}
    \zeta(\mathbf{x},t) = \delta N = \frac{\partial N}{\partial \phi_{I,i}} \delta \phi_{i} + \frac{1}{2} \frac{\partial^2N}{\partial\phi_{I,i} \partial \phi_{I,j}} \delta \phi_{i} \delta \phi_j  + \cdots, \label{zetaf}
\end{align}

\noindent where the subscripts $i$ and $j$ denote various fields and run from $1$ to $P$, and the subscript  $I$ denotes the initial value of the fields. When we consider only one curvaton field, the scalar power spectrum and bispectrum can be written as
\begin{eqnarray}\label{eq:deltaNPs}
	P_{\zeta} = \left(\frac{H_I}{2 \pi}\right)^2 \left(\frac{\partial N} {\partial \phi_I} \right)^2,\\ \label{eq:deltaNfNL}
	f_{NL} =  	\frac{5}{6} \frac{\partial^2 N}{\partial \phi_I^2} /{ \left(\frac{\partial N}{\partial \phi_I} \right)^2}.
\end{eqnarray}
The spectral index can be computed as
\begin{equation}\label{eq:ns}
	n_s - 1 \simeq \frac{2}{3} \frac{V^{\prime \prime}(\phi_I)}{H_I^2} + 2 \frac{\dot{H_I}}{H_I^2},
\end{equation}
where $V(\phi)$ is the curvaton potential and $H_I$ is the Hubble parameter when the relevant scale exits the Hubble horizon. Here we consider that the curvaton field does not evolve during inflation due to Hubble friction. The last term in Eq.~\eqref{eq:ns} is the negative of first slow-roll parameter. In this manuscript we chose the parameters such that the first term in Eq.~\eqref{eq:ns} is sub-dominant compared to second term and $n_s \simeq 0.96$ can be achieved by considering the first slow-roll parameter $-\frac{\dot{H_I}}{H_I^2} \simeq 0.02$ \cite{Gordon:2002gv}. It is  important to note that this large value of the first slow-roll parameter is consistent with our consideration that perturbations produced by inflation is negligible \cite{Gordon:2002gv,Dimopoulos:2002kt}.

The general notion of computing the derivative of the number of e-folds is to start with the following equation which denotes the energy density of the universe at the time of curvaton decay: 
\begin{equation}\label{eq:energy}
	3 M_P^2 H(t_{dec})^2 = 3 M_P^2 \Gamma^2 = \rho_{R,dec} + \rho_{\phi,dec} = \rho_{R,I} a^{-4}  +   \rho_{\phi,I} a^{-3},
\end{equation}

\noindent where $\rho_{R,dec}$ and $\rho_{\phi,dec}$ are the radiation and curvaton energy density, respectively, at the time of curvaton decay $(t_{dec})$ with $\Gamma$ being the decay width of the curvaton, ``$a$" is the scale factor of the universe, and $M_P = 2.4 \times 10^{18}$ GeV is the reduced Planck mass. Here $H$ is the post-inflationary Hubble parameter, and at the time of curvaton decay, $H(t_{dec}) \sim \Gamma$. Also, $ \rho_{R,I}$ and  $\rho_{\phi,I} $ are the radiation and curvaton energy density at the end of inflation \footnote{In our analysis we assume instantaneous  reheating of the universe right after the end of inflation.}. 
Now we take the derivative of Eq.~\eqref{eq:energy} with respect to the initial curvaton field value $\phi_I$ during (at the end of) inflation. For a quadratic potential of curvaton $V(\phi) = \frac{1}{2} m^2 \phi^2$, the derivatives of $N$ can be estimated \cite{Kawasaki:2011pd} as
\begin{eqnarray}\label{eq:ShoN}
	{\frac{\partial N}{\partial \phi_I}} &=& \frac{2 r_d}{3r_d+4} \frac{1}{\phi_I},\\ \label{eq:ShoN2}
	f_{NL} &=& {\frac{5}{12}} \left(-3 + {\frac{4}{r_d}} + {\frac{8}{4+ 3 r_d}}\right),
\end{eqnarray}

\noindent where $r_d = \frac{\rho_{\phi,dec}} {\rho_{R,dec}}$. 

Although this approach of starting with Eq.~\eqref{eq:energy} and computing the derivatives of $N$ with respect to $\phi_I$ is quite useful for the system with quadratic potential, it will often lead to wrong results for general potentials because the assumption that the energy density of the curvaton with a general potential redshifts as $a^{-3}$ is not true. In the following section, we implement an alternative method to compute the scalar power spectrum and bispectrum (for a similar approach see \cite{Kawasaki:2011pd}). 

\section{New Method for Computing Scalar Power Spectrum and Bispectrum}\label{sec:Method}

We develop an alternative approach to study the curvaton scenario for the following systems:
\begin{itemize}
	  \item Quadratic potential: $V(\phi) = \frac{1}{2}m^2 \phi^2$
	  \item Quadratic + Quartic potential: $V(\phi) =  \frac{1}{2}m^2 \phi^2 + \frac{1}{4} \lambda \phi^4$
\end{itemize}

\noindent As the curvaton with quadratic potential is studied extensively in the literature, we will check the consistency of our method to the existing results of quadratic curvaton.

\subsection{Quadratic Potential}\label{sec:SHO}
In the simplest scenario with quadratic potential, as the radiation energy density is red-shifting faster than the curvaton,  the energy density of the universe after inflation will be dominated by radiation and then will be dominated by curvaton if it lives long enough. This corresponds to $r_d \gg 1$. It is also possible that the curvaton decays very early while the universe energy density was dominated by radiation. This case corresponds to $r_d \ll 1$. During radiation domination, the scale factor behaves as $a(t) \propto t^{1/2}$, while it behaves as $a(t) \propto t^{2/3}$ during matter domination, where ``$t$" is cosmic time.  The matter-radiation equality time is estimated as
\begin{equation}\label{eq:transitionTIme}
	t_{MD} =  t_I \left(\frac{\rho_{R,I}}{\rho_{\phi,I}}\right) ^2, 
\end{equation}

where the time $t_I$ denotes the onset of curvaton oscillation when the post-inflationary Hubble parameter becomes \cite{Kawasaki:2011pd,Harigaya:2012up}
	\begin{equation}
		H_c^2 = \frac{1}{5 \phi} \frac{\partial V}{\partial \phi} = \frac{1}{5}m^2. 
\end{equation}
We choose initial conditions of the curvaton such that at the onset of post-inflationary evolution, curvaton energy density is still sub-dominant compared to radiation. As the radiation energy density redshifts as $a^{-4}$, $t_I$ can be computed as
\begin{equation}
	t_I = \sqrt{\frac{5}{4}} \frac{1}{m}.
	\end{equation}
 For $t<t_{MD}$ we solve the following sets of equation:
\begin{eqnarray}\label{SHOradEq1}
	\rho_{\phi}^{\prime}(t) &=& -\frac{3}{2t} \rho_{\phi}(t),\\ \label{SHOradEq2}
	\rho_R^{\prime}(t)  &=& -\frac{3}{2t} \left(1+\frac{1}{3}\right) \rho_{R}(t),
\end{eqnarray} 

\noindent with the initial conditions, $\rho_R(t_I) = \rho_{R,I} = 3 H_I^2$ and $\rho_{\phi}(t_I) = \rho_{\phi,I} = \frac{1}{2} m^2 \phi_I^2$, where prime denotes derivative with respect to time. The Hubble parameter in radiation dominated region $(H_R)$ is given by
\begin{equation}\label{SHOradH}
	H_{R}(t) = \left[\frac{1}{3}  \rho_{\phi}(t) + \frac{1}{3} \rho_{R}(t)\right]^{\frac{1}{2}},
\end{equation}

\noindent where we have used the Planck units $M_P = 1$.

Similarly, we solve the following equations for $t > t_{MD}$ when the curvaton starts to dominate over radiation,
\begin{eqnarray}\label{SHOmatEq}
	\rho_{\phi}^{\prime}(t) &=& -\frac{2}{t+ \frac{1}{3} t_{MD}(\phi_I)} \rho_{\phi}(t),\\
	\rho_R^{\prime}(t)  &=& -\frac{2}{t + \frac{1}{3} t_{MD}(\phi_I)} \left(1+\frac{1}{3}\right) \rho_{R}(t).
\end{eqnarray} 

\noindent Note that the term of $\frac{1}{3} t_{MD}(\phi_I)$ is introduced to regulate the Hubble behavior properly at the time of transition from radiation to curvaton dominated epoch. Considering the time dependence of the Hubble parameter at $t<t_{MD}$ and $t>t_{MD}$, $H_R = \frac{1}{2t}$ and $H_M = \frac{2}{3t}$, respectively, we see that they do not match at $t=t_{MD}$. Thus we have proposed the form,
\begin{equation}
	H_M(t) = \frac{2}{3 (t - C t_{MD})},
\end{equation}

\noindent with $C = -\frac{1}{3}$ so that $H_R(t_{MD}) = H_M(t_{MD})$ is satisfied and $H_M \simeq \frac{2}{3 t}$ for $t \gg t_{MD}$.

The evolution of Hubble parameter in the curvaton dominated epoch is given by,
\begin{equation}\label{SHOmH}
	H_{M}(t) = \left[\frac{1}{3}  \rho_{\phi}(t) + \frac{1}{3} \rho_{R}(t)\right]^{\frac{1}{2}}.
\end{equation}

\noindent With these expressions of $H_R(t)$ and $H_M(t)$, we can compute the number of e-folds $N$ from the relation $dN = H dt$ and then compute the  power spectrum and bispectrum for the scalar perturbation.

\subsubsection{Analysis Method}

 We first compute $H_R$ and $H_M$ according to Eqs.~\eqref{SHOradH} and \eqref{SHOmH} and then compute the quantity $\frac{d H_{R,M}}{d \phi_I}$. After this, $\frac{d N}{d \phi_I}$ is computed by integrating $\frac{d H_{R,M}}{d \phi_I}$ over time. 

 In the radiation dominated region,
\begin{equation}\label{eq:dndphi1}
	\frac{d N_R}{d \phi_I} = \int_{t_I}^{t_{MD}}dt~  \frac{d H_R(\phi_I,t)}{d \phi_I} + H_R(\phi_I,t_{MD}) \frac{d t_{MD}}{d \phi_I},
\end{equation}

\noindent where the second term arises because the transition time $t_{MD}$ is also a function of $\phi_I$ as it is evident from Eq.~\eqref{eq:transitionTIme}. 

 In the curvaton dominated region, we can write a similar equation,
\begin{equation}\label{eq:dndphi2}
	\frac{d N_M}{d \phi_I} = \int_{t_{MD}}^{t_f}dt ~ \frac{d H_M(\phi_I,t)}{d \phi_I} - H_M(\phi_I,t_{MD}) \frac{d t_{MD}}{d \phi_I},
\end{equation}

\noindent where the minus sign in the second term arises because the integration limit is from $t_{MD}$ to $t_f$, where $t_f$ is the curvaton lifetime.

Since $H_R(\phi_I,t_{MD}) = H_M(\phi_I,t_{MD})$ we can write, 
\begin{equation}\label{eq:ShoTotal}
	\frac{dN}{d \phi_I} = \frac{d N_R}{d \phi_I} +	\frac{d N_M}{d \phi_I} =  \int_{t_I}^{t_{MD}}dt ~ \frac{d H_R(\phi_I,t)}{d \phi_I}  +  \int_{t_{MD}}^{t_f} dt ~ \frac{d H_M(\phi_I,t)}{d \phi_I}.
\end{equation}

\begin{figure}
	\centering
	\includegraphics[width=0.8\textwidth]{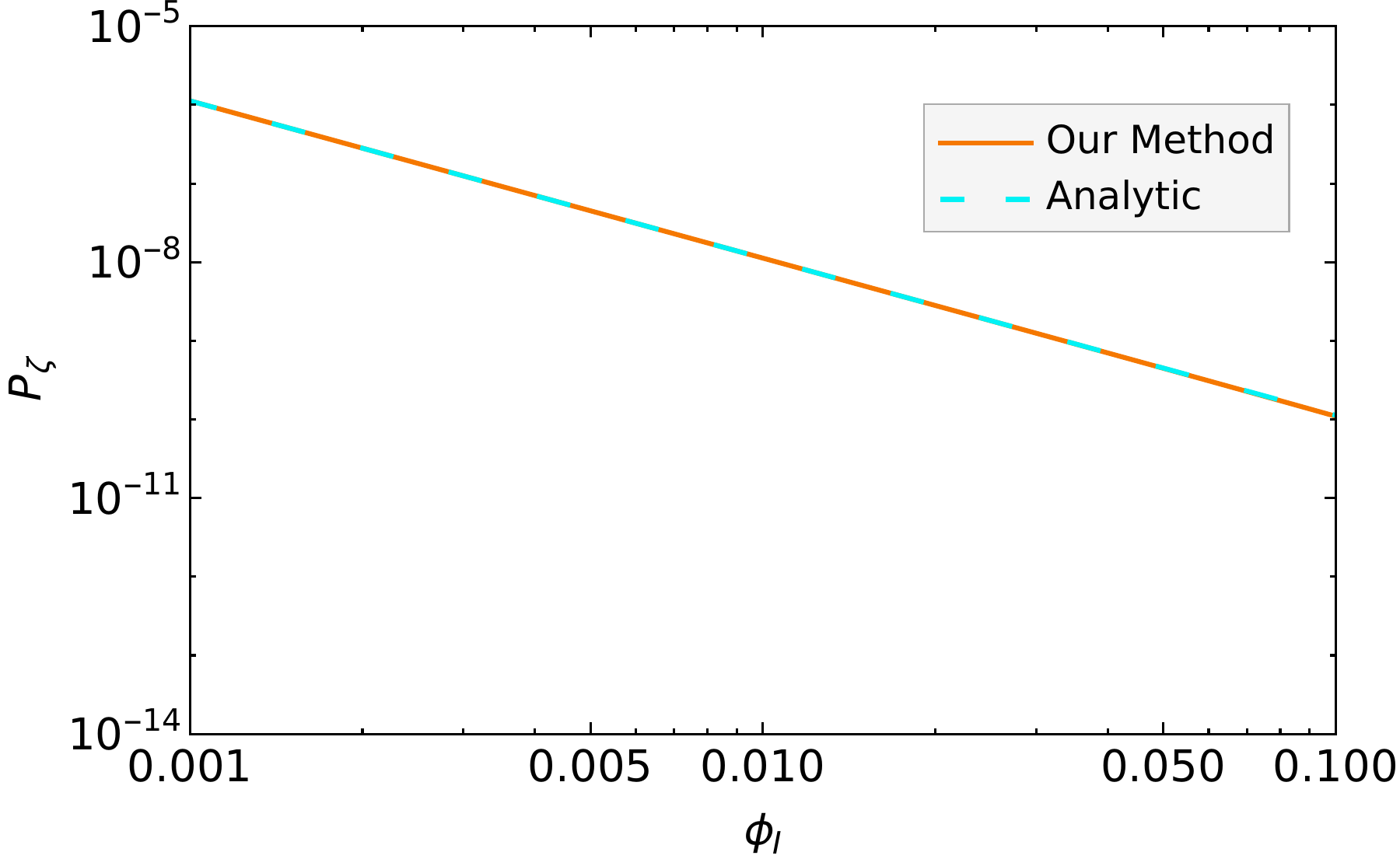}    
	\caption{\it  Comparison between scalar power spectrum estimated by our method and analytic approach. Here we have set the parameters as: $H=10^{-5},~m=10^{-6},~t_f=10^{30}$. The choice of $t_f = 10^{30}$ corresponds to $r_d \gg 1$, such that the curvaton energy density dominates the radiation energy density at the time of decay. Our result matches with the analytic estimation (solid and dashed lines are overlapping). All dimensionful quantities are expressed in the units of $M_P = 1$.}
	\label{fig:SHOPs}
\end{figure}

\begin{figure}
	\centering
	\includegraphics[width=0.8\textwidth]{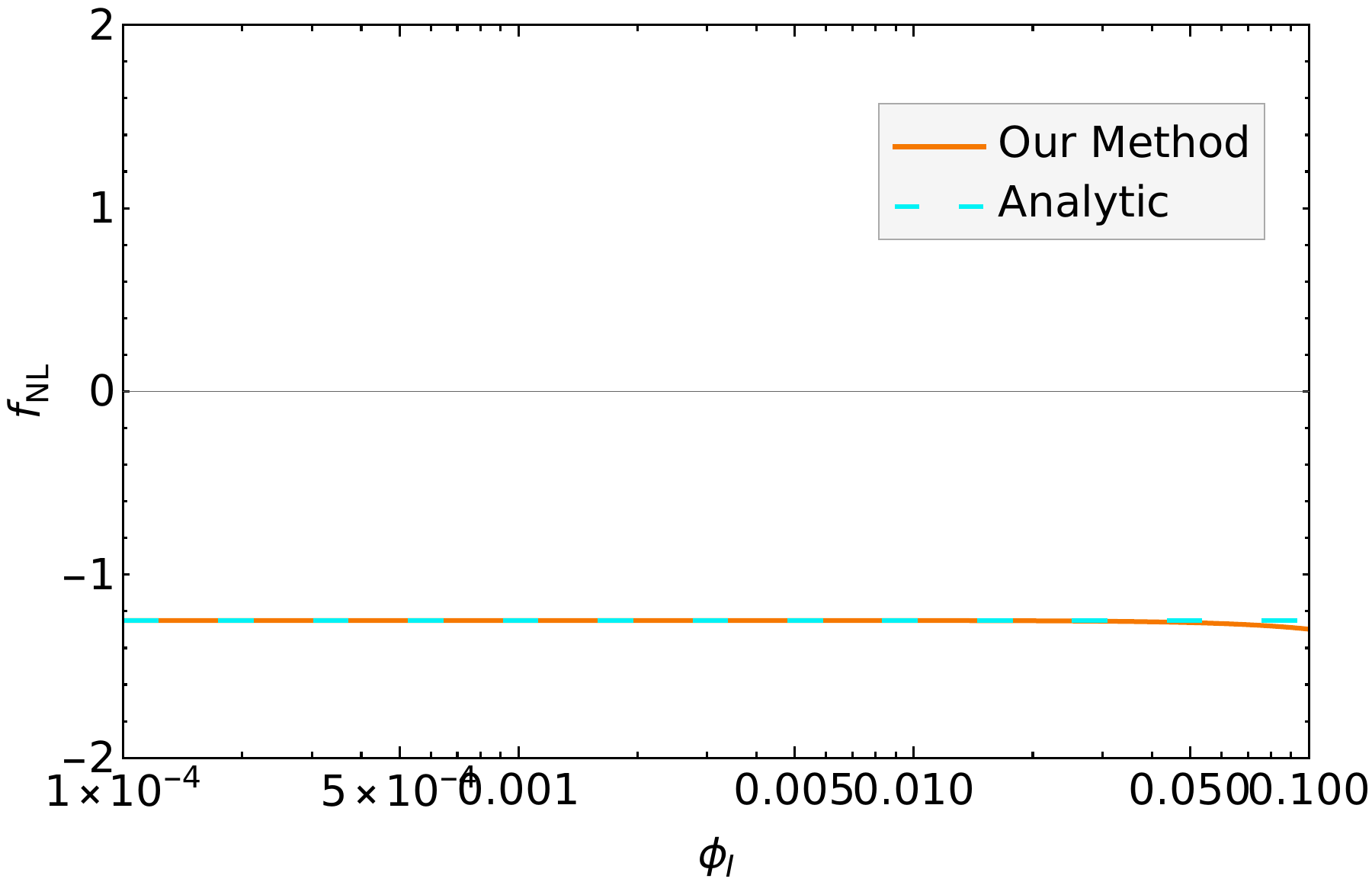}    
	\caption{\it  Comparison between scalar bispectrum estimated by our method and analytic approach. Our result matches with analytic estimation (solid and dashed lines are overlapping). The field values are expressed in the units of $M_P = 1$.}
	\label{fig:SHOfnl}
\end{figure}

\noindent We have estimated the scalar power spectrum and our result is shown in Fig.~\ref{fig:SHOPs}. We can see that our result completely matches with the well-known analytical estimation \cite{Kawasaki:2011pd}. Here we have set $t_f = 10^{30}$ which corresponds to $r_d \gg 1$, namely, the curvaton energy density dominates the radiation energy density at the time of decay. To compute the bispectrum we  take the second derivative of Eq.~\eqref{eq:ShoTotal} and use the definition Eq.~\eqref{eq:deltaNfNL}. The scalar bispectrum estimated with our method is shown in Fig.~\ref{fig:SHOfnl}. We can see that our result also matches with the standard bispectrum expression at $r_d \gg 1$ \cite{Kawasaki:2011pd}. For $r_d \ll 1$, we only compute Eq.~\eqref{eq:dndphi1} and its derivatives for power spectrum and bispectrum. 
More details about our scheme of computation and estimation of trispectrum $(g_{NL})$ will be found in \cite{CurvatonMethod}.


\subsection{Quadratic + Quartic Potential}\label{sec:quartic}

Next, we consider the following potential as the curvaton potential:
\begin{equation}\label{eq:quartic-pot}
    V(\phi) =  \frac{1}{2}m^2\phi^2 + \frac{1}{4} \lambda \phi^4.
\end{equation}
Similar curvaton potential is also analyzed in \cite{Byrnes:2014xua,Harigaya:2012up,Enqvist:2009zf} based on tracking the time when curvaton starts its oscillation in the quadratic part of the potential, but our treatment is different from these works. We first introduce what we call the ``transition field value" $(\phi_T)$ which is determined by the condition $\frac{1}{2} m^2 \phi_T^2 = \frac{1}{4} \lambda \phi_T^4$ and hence
\begin{equation}\label{eq:tranSHO}
	\phi_T = \sqrt{\frac{2}{\lambda}} m. 
\end{equation}
For $\phi > \phi_T$, the curvaton potential is dominated by the quartic term and we can approximate it as $V \simeq \frac{1}{4} \lambda \phi^4$, while for $\phi < \phi_T$ the potential is approximated as $V \simeq \frac{1}{2} m^2 \phi^2$ where the quadratic term dominates. In our computation to explore the dependence of $N$ on the initial value  of curvaton $\phi_I$, we focus on the case: $\phi_I > \phi_T$. For $\phi_I< \phi_T$ the potential is dominated by the quadratic term and we can apply the same expressions as developed in Sec.~\ref{sec:SHO}. 

For $\phi > \phi_T$, $V \simeq \frac{1}{4} \lambda \phi^4$ and the field does not immediately start evolving as the post-inflationary Hubble friction is still large to prevent curvaton evolution. As noted in \cite{Kawasaki:2011pd,Harigaya:2012up} the field will start evolving once the following condition is met
\begin{equation}\label{eq:curv-osc}
	\frac{1}{5 \phi} \frac{\partial V}{\partial \phi} = H_c^2,	
\end{equation} 
 where $H_c$ is the post-inflationary Hubble parameter at the onset of curvaton evolution. Here we consider initial condition of curvaton such that at the onset of evolution curvation is  sub-dominant in energy density, and $H_c$ will be solely determined by radiation. After Eq.~\eqref{eq:curv-osc} is satisfied curvaton  will start evolving as $\phi \propto a^{-1}$.  As radiation energy density evolves as $a^{-4}$ we can compute the time when curvaton starts evolving using Eq.~\eqref{eq:curv-osc} as
 \begin{equation}
 	t_I = \sqrt{\frac{5}{4\lambda}} \frac{1}{\phi_I}.
 \end{equation}
 We can track the time when $\phi$ becomes equal to $\phi_T$,
\begin{equation}\label{eq:fieldTime}
	t_T = t_I \left(\frac{\phi_I}{\phi_T}\right)^2 = t_I  \frac{\lambda}{2} \left(\frac{\phi_I}{m}\right)^2.
\end{equation}

\noindent After this time the quadratic potential will dominate, and we apply the analysis of the previous section. Again we can first determine the Hubble parameter in different stages and then we take derivative with respect to $\phi_I$, and integrate it over time with appropriate limits to determine $\frac{dN}{d\phi_I}$. 

We express $\frac{dN}{d\phi_I}$ for three separate regions as follows:
\begin{eqnarray}\label{eq:dNquartic1}
	\frac{d N_{R,1}}{d\phi_I} &=& \int_{t_I}^{t_T} dt ~ \frac{d H_{R,1}(\phi_I,t)}{d \phi_I} + H_{R,1}(\phi_I,t_T) \frac{d t_T}{d \phi_I},\\ \label{eq:dNquartic2}
	\frac{d N_{R,2}}{d\phi_I} &=& \int_{t_T}^{t_{MD}} dt ~ \frac{d H_{R,2}(\phi_I,t)}{d \phi_I} + H_{R,2}(\phi_I,t_{MD}) \frac{d t_{MD}}{d \phi_I} - H_{R,2}(\phi_I,t_T) \frac{d t_T}{d \phi_I}, \\ \label{eq:dNquartic3}
	\frac{d N_{M}}{d\phi_I} &=& \int_{t_{MD}}^{t_f} dt ~ \frac{d H_M(\phi_I,t)}{d \phi_I} - H_M(\phi_I,t_{MD}) \frac{d t_{MD}}{d \phi_I}.
\end{eqnarray}

\noindent Again note that the boundary terms involving $ \frac{d t_{T,MD}}{d \phi_I}$'s cancel each other and only the integral parts contribute to the generation of curvature perturbation. 
In this paper, we will be only interested in $r_d \gg 1$. 

\begin{figure}
	\centering
	\includegraphics[width=0.8\textwidth]{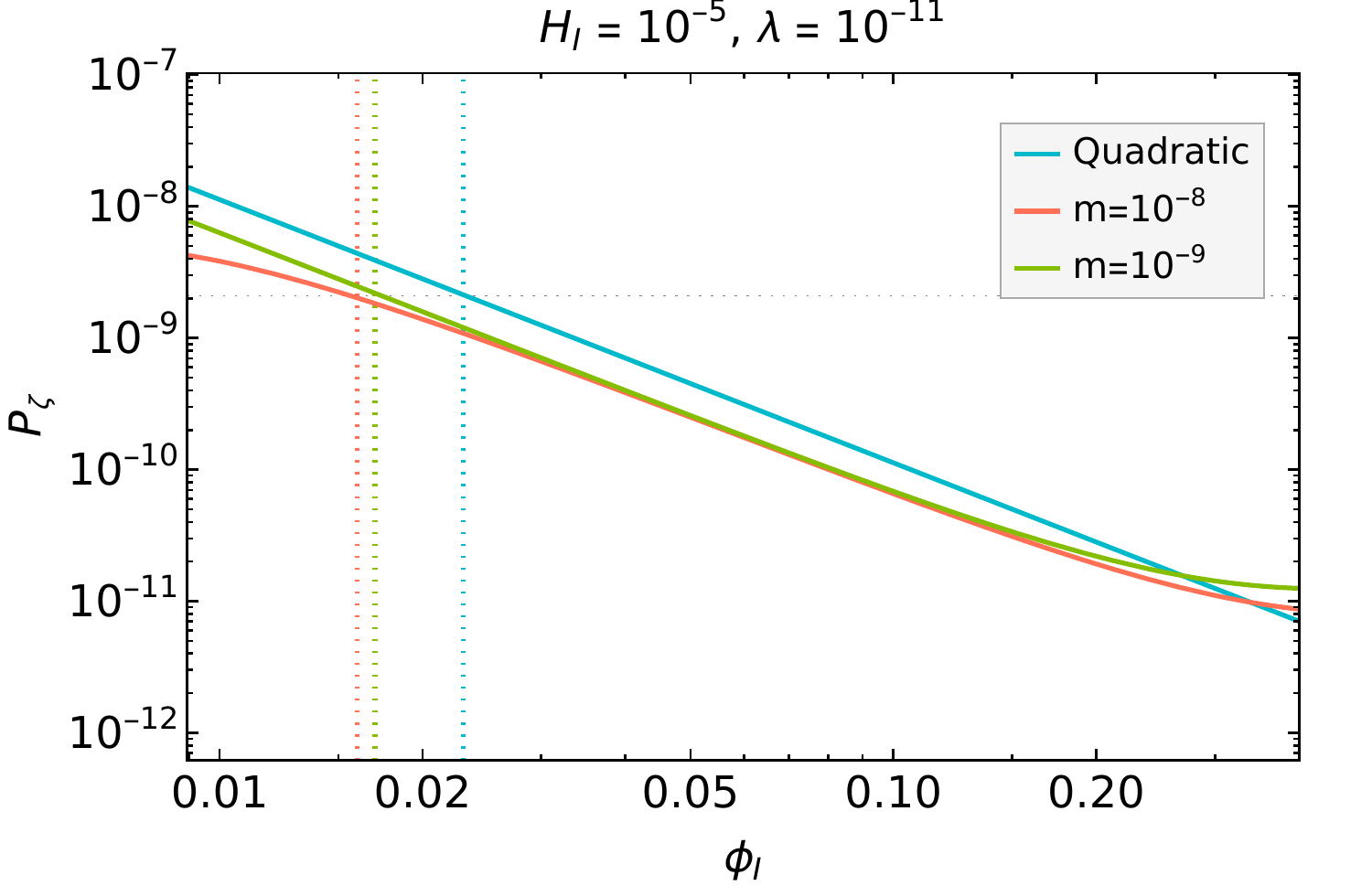}    
	\caption{\it Plot for the power spectrum's dependence on $\phi_I$ with different choices of mass while $H_I= 10^{-5}$, $\lambda = 10^{-11}$. The horizontal dotted black line denotes the Planck normalization for power spectrum, and the vertical dotted lines denotes the $\phi_I$ values required to meet this value for different choices of mass. In these plots curvaton dominates the energy density before decay. The dimensionful parameters are expressed in the units of $M_P = 1$. }
	\label{fig:quarticPsm}
\end{figure}

\begin{figure}
	\centering
	\includegraphics[width=0.8\textwidth]{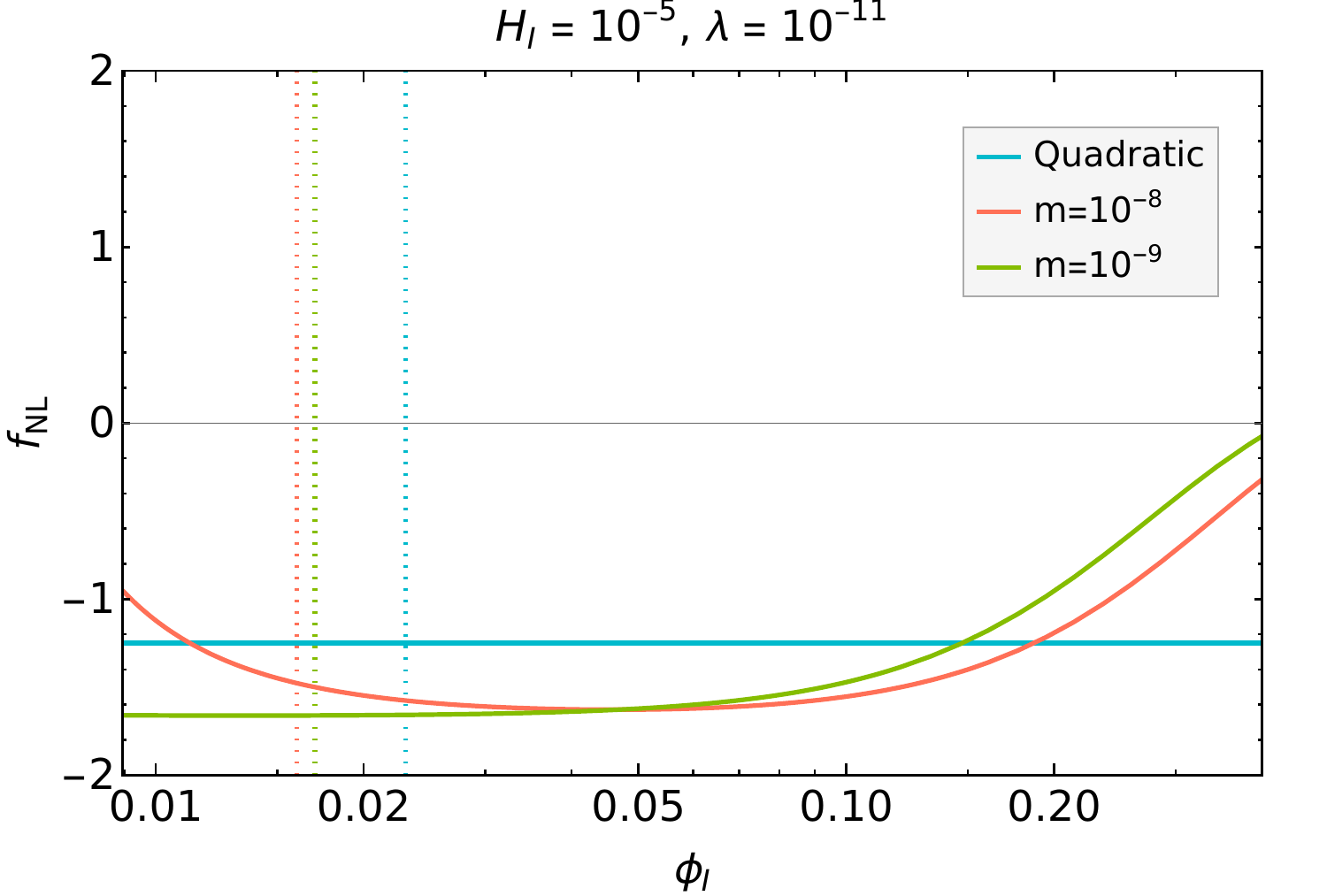}    
	\caption{\it Plot for the bispectrum's dependence on $\phi_I$ with different choices of mass while $H_I= 10^{-5}$, $\lambda = 10^{-11}$. The vertical dotted lines denotes the required value of $\phi_I$ to match Planck normalization of power spectrum. In these plots curvaton dominates the energy density before decay. The dimensionful parameters are expressed in the units of $M_P = 1$.}
	\label{fig:quarticfNLm}
\end{figure}
In Figs.~\ref{fig:quarticPsm} and \ref{fig:quarticfNLm} we plot the dependence of power spectrum and bispectrum, respectively, on the initial curvaton field value. Here, the curvaton decay time $t_f$ in Eq.~\eqref{eq:dNquartic3} chosen such that $r_d \gg 1$ for all the initial curvaton field value $\phi_I$. From  Figs.~\ref{fig:quarticPsm} and \ref{fig:quarticfNLm} we can see that the dependence of both the power spectrum and bispectrum for ``Quadratic + Quartic" scenario is different from the simple quadratic scenario. Especially from Fig.~\ref{fig:quarticfNLm} we can see that $\vert f_{NL} \vert$ can be larger than the quadratic case for suitable choice of $\phi_I$.


\section{CMB Observables with Late Curvaton Decay}\label{sec:smallR}

\begin{figure}
	\centering
	\includegraphics[width=0.8\textwidth]{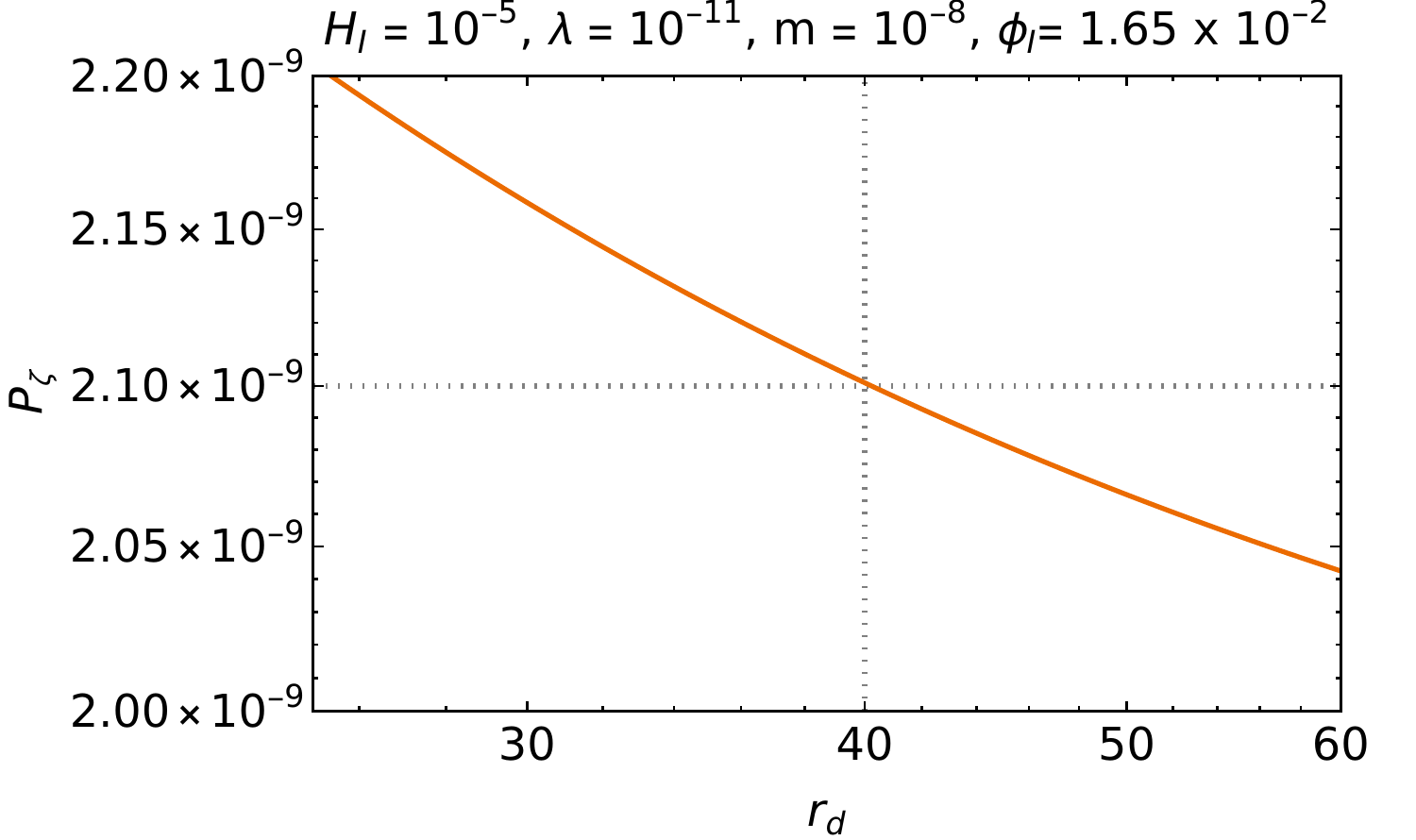}    
	\caption{\it Plot for the power spectrum's dependence on $r_d$ while $m = 10^{-3}$, $\lambda = 10^{-11}$, $\phi_I = 3.5 \times 10^{-2}$. The dimensionful parameters are expressed in the units of $M_P = 1$. The power spectrum $P_{\zeta} = 2.1\times 10^{-9}$ can be achieved at $r_d = 40$.}
	\label{fig:psVrphiT}
\end{figure}

\begin{figure}
	\centering
	\includegraphics[width=0.8\textwidth]{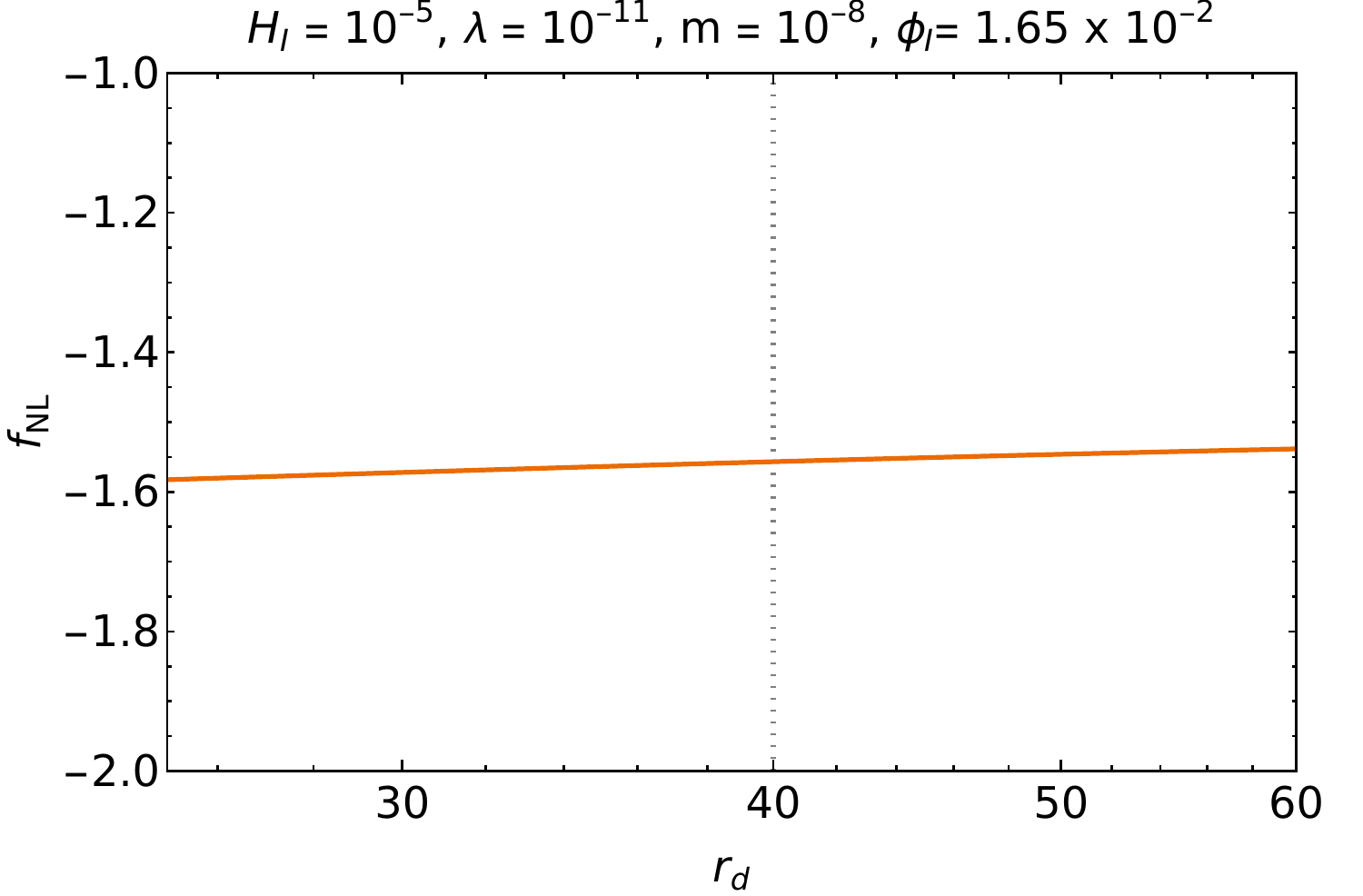}    
	\caption{\it Plot for the dependence of the amplitude of bispectrum on $r_d$  while $m = 10^{-8}$, $\lambda = 10^{-11}$, $\phi_I = 1.65 \times 10^{-2}$. The dimensionful parameters are expressed in the units of $M_P = 1$. Amplitude of bispectrum $f_{NL} = -1.56$ at $r_d = 40$.}
	\label{fig:fnlVrphiT}
\end{figure}

 Let us assume that the observed scalar power spectrum originates dominantly from the curvaton. Curvaton scenario can generate observable scalar powerspectrum both for early decay $(r_d \leq 1)$ and late decay $(r_d \gg 1)$ \cite{Byrnes:2014xua}. In this article one of our main interests is to realize baryogenesis from the curvaton scenario, namely, the curvaton itself is responsible for the observed baryon asymmetry along with producing the observable scalar curvature perturbation (see Sec.~\ref{sec:ADbaryo}). It is well known that correlated or anti-correlated baryon (dark matter) isocurvature perturbation can be generated if baryon asymmetry (dark matter) is produced from curvaton decay \cite{Lyth:2003ip,Lemoine:2006sc,Lemoine:2008qj}. Current observation suggests that correlated baryon isocurvature perturbation $\mathcal{S}_B \sim 0$ \cite{Planck:2018vyg}, and in a curvaton scenario this limit can be achieved if curvaton dominates the energy density of the universe before its decay (see Appendix~\ref{app:iso}). For a benchmark parameter choice $\lbrace H_I = 10^{-5}, ~m= 10^{-8}, ~\lambda = 10^{-11}, ~\phi_I = 1.65\times 10^{-2} \rbrace$, we show the dependence of scalar power spectrum and bispectrum on the parameter $r_d$ in Figs.~\ref{fig:psVrphiT} and \ref{fig:fnlVrphiT}. In Fig.~\ref{fig:psVrphiT} one can see that Planck normalized power spectrum is reproduced by $r_d \sim 40$ and the corresponding amplitude of bispectrum is found to be $f_{NL}\sim -1.56$ in Fig.~\ref{fig:fnlVrphiT},  which is well within the constraint of Planck non-Gaussianity bound\footnote{Although we can evaluate amplitude of the trispectrum $g_{NL}$ with our method, we skip evaluating this observable as it is very loosely constrained by current observations \cite{Planck:2018vyg}. This analysis will be shown in \cite{CurvatonMethod}.}. Here, the following definition for the amplitude of scalar bispectrum has been used \cite{Huang:2008rj}: 
\begin{equation}\label{eq:fnlNorm}
    f_{NL} = \beta^2 \frac{5}{6} \frac{d^2 N}{d \phi_I^2}/\left(\frac{d N}{d \phi_I}\right)^2,
\end{equation}
where $\beta = \frac{P_{\zeta,curvaton}}{P_{\zeta,Total}}$. $P_{\zeta,curvaton}$ is same as Eq.~\eqref{eq:deltaNPs}, and $P_{\zeta,Total} = 2.1 \times 10^{-9}$ is the Planck's normalization of the scalar power spectrum \cite{Planck:2018vyg}. 
Note that $f_{NL}$ is normalized by $\beta$ since, in general, the curvaton's contribution to the Planck normalized power spectrum is not necessarily dominating. In this case, $f_{NL}$ evaluated by Eq.~\eqref{eq:deltaNfNL} is not the correct value. 
 
\begin{figure}
	\centering
	\includegraphics[width=0.8\textwidth]{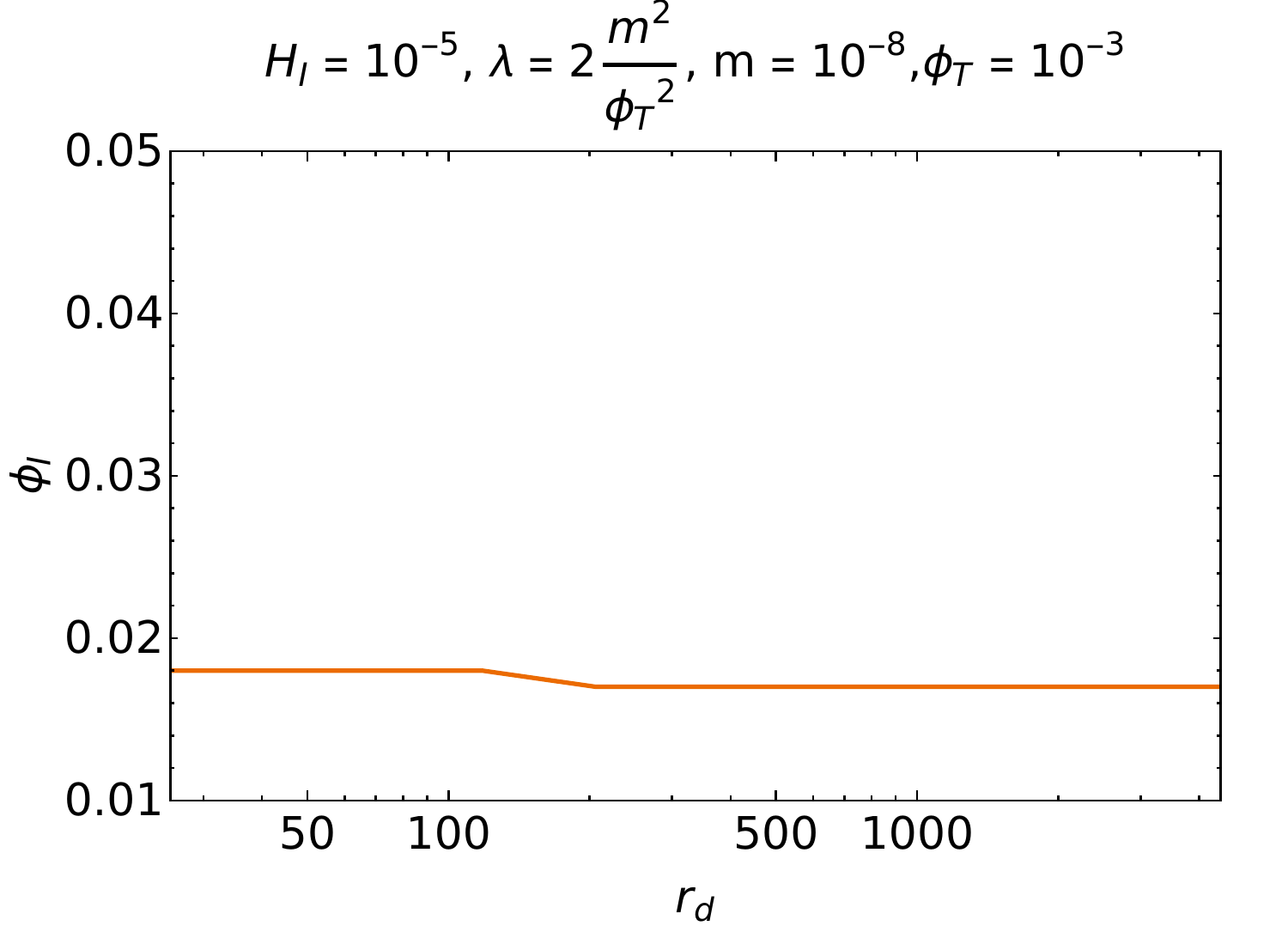}    
	\caption{\it 
Relation between $\phi_I$ and $r_d$ with the other parameters fixed as indicated. 
Along the line, the Planck normalized power spectrum is reproduced only by the curvaton. The dimensionful parameters are expressed in the units of $M_P = 1$.
}
	\label{fig:phiVr}
\end{figure}
In the subsequent analysis with the curvaton potential of Eq.~\eqref{eq:quartic-pot}, we treat the transition field value $\phi_T$ as a free parameter. The region of our interest will be $r_d \gg 1$. 
As mentioned above, we treat $\phi_T$ as a free parameter in our analysis and hence the quartic coupling is determined as
\begin{equation}
	\lambda = 2 \frac{m^2}{\phi_T^2},
\end{equation}
\noindent once another free parameter $m$ is fixed. For various values of $\phi_I$ with fixed values of $H_I$, $m$ and $\phi_T$, we calculate the scalar power spectrum 
and determine $r_d$ to reproduce the Planck normalized value. 
Our result for the relation between $\phi_I$ and $r_d$ is shown in Fig.~\ref{fig:phiVr}. Here, we considered a region where $\phi_I > \phi_T$ such that the curvaton evolves through the quartic region before transitioning to the quadratic region. We can see that $\phi_I$ is saturated to be almost independent of $r_d$ for $r_d \gg 1$. 

\section{Affleck-Dine Curvaton and Baryogenesis}\label{sec:ADbaryo}

To explore baryogenesis produced from a curvaton scenario, we consider the curvaton  as an Affleck-Dine (AD) field with a potential of the form \cite{Affleck:1984fy},
\begin{equation}\label{eq:AD-pot}
    V(\Phi) = m^2 (\Phi^{\dagger} \Phi) + \lambda (\Phi^{\dagger} \Phi)^2 + \epsilon m^2 \left[\Phi^2 + (\Phi^{\dagger})^2\right],
\end{equation}

\noindent where the AD field $\Phi$ carries a baryon/lepton number $(Q_{\phi})$, and the last term proportional to $\epsilon$ explicitly violates the baryon/lepton symmetry. For this scenario, we consider the following points:

\begin{itemize}
    \item The $\Phi$ field can be written as
    \begin{eqnarray}
     \label{eq:adRadial}
           \Phi &=& \frac{1}{\sqrt{2}}\phi_r e^{i \theta}\\
            \label{eq:adComplex}
         &=& \frac{1}{\sqrt{2}}\left(\phi_1 + i \phi_2\right),
    \end{eqnarray}
    
    where $\phi_r$ is the radial part, and $\theta$ is the angular part, and $\phi_1$ and $\phi_2$ are the real and imaginary components of $\Phi$, respectively.
    \item We consider $0< \epsilon \ll 1$ such that the trajectory of the AD field evolution is almost a straight line.
    \item Since $\epsilon \ll 1$ we can see that the last term of Eq.~\eqref{eq:AD-pot} contributes negligibly to the energy density of the universe. The kinetic part of the AD field in terms of the radial and angular part can be written as \footnote{In $(-,+,+,+)$ signature.}
    \begin{equation}\label{eq:AD-kin}
        K_{\Phi} = -\frac{1}{2}\partial_{\mu} \phi_r \partial^{\mu} \phi_r - \frac{1}{2}\phi_r^2 \partial_{\mu} \theta \partial^{\mu} \theta,
    \end{equation}
     and we can write the equation of motion to be
     \begin{eqnarray}\label{eq:adRadEom}
        \ddot{\phi}_r + \phi_r \dot{\theta}^2 + 3 H \dot{\phi}_r + m^2 \phi_r +\lambda \phi_r^3 + 2\epsilon m^2 \phi_r \cos{2 \theta} &=& 0, \\ \label{eq:adAngEom}
        \phi_r\ddot{\theta} + 3 H \phi_r \dot{\theta} + 2 \dot{\phi_r} \dot{\theta} - 2\epsilon m^2 \phi_r \sin{2 \theta} &=& 0,
    \end{eqnarray}
     where $H$ is the post-inflationary Hubble parameter. The energy density of the system is given by
    \begin{equation} \label{EM00}
        \rho_{\Phi} = \frac{\dot{\phi}_r^2}{2} + \frac{\phi_r^2 \dot{\theta}^2}{2} + \frac{m^2}{2} \phi_r^2 + \frac{\lambda}{4} \phi_r^4 + \epsilon m^2 \phi_r^2 \cos{2 \theta}.
    \end{equation}
    In Eq.~\eqref{eq:adRadEom} we can see that for $\epsilon \ll 1$ the radial field $\phi_r$ evolves independently of $\theta$ if $\dot{\theta}^2 \ll m^2$. From Eq.~\eqref{eq:adAngEom} we can see that for $\epsilon \ll 1$, $\theta \simeq \text{constant}$ is a solution so that  $\dot{\theta} = 0$. For $0< \epsilon \ll 1$ we expect $\dot{\theta} \sim 0$ and this ensures the radial field evolves almost independently of the angular field. From  Eq.~\eqref{EM00} we can see that terms involving $\theta$ and $\dot{\theta}$ are negligibly small. Thus, $H$ is controlled by $\phi_r$ and almost independent of $\theta$, and as a result, the number of e-folds $N$ in post-inflationary scenario is mainly controlled by $\phi_r$. We then identify $\phi_r$ as the curvaton with the potential approximately to Eq.~\eqref{eq:quartic-pot} which is responsible for the production of density perturbation. As we will discuss below, the $\epsilon$ parameter plays a crucial role to produce the observed baryon asymmetry.
\end{itemize}

Let us now briefly discuss the evolution of the AD field in the light of Sec.~\ref{sec:quartic} and the generation mechanism of the baryon asymmetry. We have already identified the radial direction $\phi_r$ as the curvaton with its potential of Eq.~\eqref{eq:quartic-pot} as a good approximation. Here we can also define a transition field value $\phi_{r,T}$ similar to the transition field value from quartic to quadratic region as Eq.~\eqref{eq:tranSHO}, and set the initial field value $\phi_{r,I} > \phi_{r,T}$.  During the evolution from $\phi_r > \phi_{r,T}$, the scalar potential is dominated by the quartic term $(\Phi^{\dagger} \Phi)^2$, which is completely symmetric and no baryogenesis takes place. When $\phi_r$ becomes less than $\phi_{r,T}$, the quadratic term starts to dominate the potential with a slight difference between $\phi_1$ and $\phi_2$ masses caused by the $\epsilon$ parameter. The evolution of the AD field with the transition from quartic to quadratic dominated potential is analyzed by the so-called threshold approximation \cite{Lloyd-Stubbs:2020sed,Babichev:2018sia}. Due to their mass splitting, the  fields $\phi_1$ and $\phi_2$  evolve with slightly different frequencies, and as a result baryon/lepton asymmetry is generated and stored as a difference between the number densities of $\Phi$ and $\Phi^{\dagger}$ \cite{Lloyd-Stubbs:2020sed}.

Following \cite{Lloyd-Stubbs:2020sed}, generated baryon asymmetry in co-moving frame is estimated as,
\begin{equation} \label{Eqn518}
    N_B = 2 Q_{\phi} m^2 \epsilon \frac{\phi_{r,I}^3}{\phi_{r,T}}  \sin(2 \theta_I) J,
\end{equation}
\noindent where $Q_{\phi}$ is the baryon/lepton number of the AD field $\Phi$, $\theta_I$ is the initial value of the angular component, and $J$ can be written as \cite{Mohapatra:2021aig}
\begin{eqnarray}
    J &=& \frac{\Gamma_{\phi}}{8 \epsilon^2 m^2} ~~ \text{for $\epsilon \gg \frac{\Gamma_{\phi}}{m}$,}\\
    &=& \frac{1}{2 \Gamma_{\phi}} ~~~~~ \text{for $\epsilon \ll \frac{\Gamma_{\phi}}{m}$,}
\end{eqnarray}

\noindent with $\Gamma_{\phi}$ being the decay width of the AD field $\Phi$. Since $\epsilon \ll 1$ this decay width is nothing but the curvaton decay width. The temperature of the universe $(T_{dec})$ when the curvaton decays is estimated as,
\begin{equation}
    \Gamma_{\phi} = \frac{3}{2} H (T_{dec}) = \frac{3}{2} \sqrt{\frac{\pi^2}{90}g_{\star}} \frac{T_{dec}^2}{M_P},
\end{equation}

\noindent where $g_{\star}$ is the number of relativistic degrees of freedom. Note that the curvaton decay reheats the universe. As the curvaton dominates the energy density before its decay and reheats the universe again, the temperature $T_{dec}$ can also be identified as the reheating temperature of the curvaton. Now, according to the discussion in Sec.~\ref{sec:quartic}, from the end of inflation to the transition point $\phi_{r,T}$, the field evolves as $a^{-1}$ and we obtain the following relation,
\begin{equation}\label{eq:field-temp}
    \frac{\phi_{r,T}}{\phi_{r,I}} = \frac{a_I}{a_T} = \frac{T_T}{T_I},
\end{equation}

\noindent where, $a_I$, $a_T$, $T_I$, $T_T$ are the scale factor at the end of inflation, scale factor at the transition point, temperature at the end of inflation and temperature at $\phi_r = \phi_{r,T}$, respectively. The transition field value $\phi_{r,T}$ can be estimated according to Eq.~\eqref{eq:tranSHO}. The temperature $T_{I}$ is related to the Hubble during inflation as
\begin{equation}\label{eq:TI-HI}
    T_I = \left(\frac{90}{\pi^2 g_{\star}}\right)^{1/4} \sqrt{H_I M_P}.
\end{equation}
\noindent Considering the evolution of the curvaton energy density as $\rho_{\phi,dec} = \rho_{\phi,I} \left(\frac{a_I}{a_T}\right)^4 \left(\frac{a_T}{a_{dec}}\right)^3$ and the radiation energy density as $\rho_{R,dec} = \rho_{R,I}\left(\frac{a_I}{a_{dec}}\right)^4$, for sufficient curvaton domination, we find
\begin{equation}
    T_{dec} = r_d^{\frac{1}{4}}T_{R,I}\frac{\phi_{r,T}}{\phi_{r,I}}\left(\frac{\phi_{r,dec}}{\phi_{r,T}}\right)^{\frac{2}{3}},
\end{equation}
\noindent where we have used $\phi_r \propto a^{-\frac{3}{2}}$ for $\phi_r \leq \phi_{r,T}$, and $\phi_{r,dec}$ is the curvaton field value at the time of decay, which can be computed as
\begin{equation}
    \phi_{r,dec} = \left\{\frac{m^2}{3H_I^2 M_P^2 r_d} \left(\frac{\phi_{r,I}}{\phi_{r,T}}\right)^{4}\right\}^{\frac{3}{2}} \phi_{r,T}^4.
\end{equation}
Here, we have used $\rho_{\phi,dec} = m^2 \phi_{r,dec}^2 = \frac{\pi^2}{30} g_{\star} T_{dec}^4$. 
The baryon-to-entropy ratio $\left(\frac{n_B}{s}\right)$ is calculated by
\begin{eqnarray} \label{eq:density}
    n_B &=& N_B \left(\frac{a_I}{a_{dec}}\right)^3 = N_B \left(\frac{a_I}{a_T}\right)^3 \left(\frac{a_T}{a_{dec}}\right)^3 = N_B \left(\frac{\phi_{r,T}}{\phi_{r,I}}\right)^3 \left(\frac{\phi_{r,dec}}{\phi_{r,T}}\right)^2, \\
    \label{eq:entropy}
    s &=& \frac{2 \pi^2}{45} g_{\star} T_{dec}^3,\\ \label{eq:nb-s}
    \frac{n_B}{s} &=& \frac{9 Q_{\phi}}{16}\sqrt{\frac{\pi^2}{90}g_{\star}}\sin(2\theta_I) \frac{T_{dec}^3}{\epsilon m_{\phi}^2 M_P}.
\end{eqnarray}
\noindent Thus by using the above formulas, $\frac{n_B}{s}$ is determined by the following set of free parameters in our analysis
\begin{equation}
    \left\lbrace H_I, \phi_I, m, \lambda, r_d, \epsilon, \theta_I, Q_{\phi} \right\rbrace .
\end{equation}
\noindent In our analysis, we set $Q_{\phi} \sin(2 \theta_I) \sim 1$ for simplicity and considered the case with $\frac{\Gamma_{\phi}}{m} \ll \epsilon \ll 1$. 

\begin{table}
	\begin{tabular}{|l|l|l|l|l|l|}
		\hline
		Benchmarks                                                                           & $f_{NL}$ & $\epsilon$            & $\frac{\Gamma_{\phi}}{m}$     & $T_{dec}$             \\ \hline
		\makecell{$H_I = 10^{-5}, m = 10^{-7}$,\\ $\phi_T = 10^{-3}, \phi_I = 1.17 \times 10^{-2}$, \\ $r_d = 278.11$}            &  $-1.66$   & $2.64 \times 10^{-11}$    & $4.27 \times 10^{-16}$   & $2.93 \times 10^{-12}$  \\ \hline
		\makecell{$H_I = 3.21 \times10^{-6}, m = 10^{-7},$ \\$ \phi_T = 10^{-3}, \phi_I = 5.54 \times 10^{-3}$,\\ $r_d = 48.74$} & $-1.64$   & $1.34 \times 10^{-10}$ & $1.28 \times 10^{-15}$  & $5.09 \times 10^{-12}$ \\ \hline
	\end{tabular}
	\caption{The $r_d$ values are chosen where the scalar power spectrum $P_{\zeta}$ matches the Planck 2018 normalization. The choice of $H_I = 3.21 \times 10^{-6}$  corresponds to the tensor-to-scalar ratio of $r = 10^{-3}$, while the choice of $H = 10^{-5}$ corresponds to $r = 9.65 \times 10^{-3}$. Here we have considered baryon-to-entropy ratio $\frac{n_B}{s} \simeq 10^{-10}$ and estimated the parameters. The $f_{NL}$ values shown here are consistent with the current Planck data and testable in future observations such as CMB-S4 \cite{CMB-S4:2022ght}, LiteBIRD \cite{LiteBIRD:2024twk}, LSS \cite{Scoccimarro:2003wn,Cabass:2022ymb,Biagetti:2021tua}, and 21-cm experiments \cite{Yamauchi:2022fri,Jolicoeur:2023tcu,Kopana:2024qqq}. All dimensionful parameters are expressed in the units of $M_P = 1$.}
 \label{Table:nb-s}
\end{table}

As discussed in Sec.~\ref{sec:quartic}, for given values of the set of parameters $\lbrace H_I, \phi_I, m,  \lambda \rbrace$, the $r_d$ value is fixed so as to reproduce the Planck normalized power spectrum, and then $f_{NL}$ is predicted. With the fixed $r_d$, we find $\epsilon$ to reproduce the observed baryon asymmetry, $\frac{n_B}{s} \simeq 10^{-10}$. We check the consistency of our analysis, $\epsilon \gg \frac{\Gamma_{\phi}}{m}$, for the resultant $\epsilon$ value. In Table \ref{Table:nb-s} we show our results for two benchmark points. 

In our scenario as baryon asymmetry is generated from the curvaton and later curvaton is decaying into baryons and anti-baryons, it is possible that there can be non-zero baryon isocurvature perturbations in tension with the current observation.  We have identified the radial part as the curvaton and sufficient domination of the curvaton along with a proper choice of the initial value of the angular part can make the isocurvature perturbation negligible as discussed in detail in Appendix~\ref{app:iso}.
 Here we also consider that dark matter is produced after curvaton decay and hence dark matter isocurvature perturbation $\mathcal{S}_{DM} \sim 0$.

\section{Conclusions and Discussions}

We have introduced a novel scenario in which the Affleck-Dine (AD) mechanism and the curvaton scenario can be described in a unified way. In this construction, the radial component of the AD field acts as a curvaton, generating the observed scalar perturbations, while the evolution of the angular component is crucial for generating the observed baryon asymmetry of the universe. 

We have first introduced a novel analytic method based on $\delta N$ formalism to compute the perturbations, which allows us to deal with the evolution of curvaton with a polynomial potential more accurately than the existing methods. 
For a given curvaton potential, the curvaton evolution is separated into distinct regions in which a single term in the curvaton potential dominates the potential energy. 
Solving the evolution of the energy density of the curvaton analytically for each region by consistently connecting transitions from one region to another, we have obtained the Hubble parameter for the entire regions of the curvaton evolution. 
Then, we systematically computed the derivatives of the number of e-folds $N$ to estimate the scalar power spectrum and bispectrum based on the $\delta N$ formalism.
We have shown that our approach reproduces the existing results for the quadratic curvaton potential in the literature \cite{Kawasaki:2011pd}.

To describe the unification of the AD field $(\Phi)$ and the curvaton, we have investigated the AD field potential consisting of quadratic and quartic terms with a small baryon/lepton-number-violating mass term. 
Since the baryon/lepton-number-violating mass term is very small compared to the symmetric mass term, 
we found that the radial component of the AD field $(\phi_r)$ evolves almost independently of the angular component  $(\theta)$, and hence the contribution of $\Phi$ to the post-inflationary Hubble parameter mainly comes from $\phi_r$. 
This allows us to identify $\phi_r$ as the curvaton while the evolution of $\theta$ is responsible for creating the difference in number densities between $\Phi$ and $\Phi^{\dagger}$ and hence, baryon asymmetry. 
We employ our analytic method to compute scalar curvature perturbations with this system. 
For $r_d \gg 1$ such that the curvaton energy density dominates over the radiation energy density before it's decay ensures vanishing of the baryon isocurvature perturbation.
We have shown two benchmark points in Table~\ref{Table:nb-s}, which simultaneously lead to the scalar power spectrum consistent with the Planck observations and the observed baryon asymmetry.
Additionally, we have found that the scalar bispectrum is negative, with an amplitude of $\vert f_{NL}\vert \sim 0.9$, 
which is consistent with the current Planck data and testable in future observations such as CMB-S4 \cite{CMB-S4:2022ght}, LiteBIRD \cite{LiteBIRD:2024twk}, LSS \cite{Scoccimarro:2003wn,Cabass:2022ymb,Biagetti:2021tua}, and 21-cm experiments \cite{Yamauchi:2022fri,Jolicoeur:2023tcu,Kopana:2024qqq}. 

Finally, to complete our scenario, we discuss how the AD/curvaton field connects to the Standard Model (SM) 
and transfers the generated baryon/lepton asymmetry to the SM sector. 
Among various possibilities, we find that the unified model for inflation, neutrino mass, dark matter, and baryogenesis proposed in \cite{Mohapatra:2021ozu} aligns well with our scenario. 
In this model, the inflaton is identified as the AD field, which carries a unit lepton number, and its evolution is governed 
by the same potential as in Eq.~\eqref{eq:AD-pot}. 
Consequently, the AD mechanism for generating lepton asymmetry in that work is fundamentally similar to the mechanism discussed in this paper.
The model incorporates a lepton-number-conserving coupling of the AD field with a right-handed neutrino and a new SM singlet fermion with a vanishing lepton number. The lepton asymmetry stored in the AD field is transferred to the SM sector through its decay into these fermions. 
In \cite{Mohapatra:2021ozu}, tiny neutrino masses are generated through the inverse seesaw mechanism, where the vacuum expectation value (VEV) of the curvaton plays a crucial role. A pseudo-Goldstone boson emerges as a dark matter candidate at low energies, involving $\phi_2$ as a component, with its mass proportional to $\epsilon$. 
Thus, the AD field $\Phi$ plays a pivotal role in both neutrino mass generation and dark matter physics, 
  providing a unified framework for the curvaton, neutrino mass, dark matter, and baryogenesis.
This dark matter interacts extremely weakly with SM particles and has never been in thermal equilibrium. 
In the early universe, this ``freeze-in" dark matter \cite{Hall:2009bx} is predominantly produced 
 through pair annihilation of SM Higgs bosons in the thermal plasma. 
 The majority of the present dark matter population is generated at $T \sim m_H$, 
 where $m_H=125$ GeV is the SM Higgs boson mass.
In our framework, the inflaton is replaced by the curvaton, with inflation driven by a separate scalar field. 
Nonetheless, the observed scalar power spectrum is entirely generated by the curvaton. 
We predict $\vert f_{NL} \vert \sim 1.6$, 
consistent with current Planck data and testable in future observations.

\acknowledgments
The work of N.O. is supported in part by the U.S. DOE Grant Nos. DE-SC0012447 and DE-SC0023713.

\appendix

\section{Analysis of Isocurvature Perturbations}\label{app:iso}
In our analysis of baryogenesis from AD curvaton field the radial part of the AD field $\phi_r$ is identified as the curvaton and it dominates the energy density before decay, while the angular part $\theta$ remains sub-dominant. Moreover as argued in Sec.~\ref{sec:ADbaryo} the angular part is almost decoupled from the radial part, thus the evolution of $\phi_r$ is almost independent of $\theta$. As $\phi_r$ is responsible for both curvature perturbation and baryogenesis, first we will compute the contribution of the radial part to the baryon isocurvature perturbation. As energy density of $\theta$ is negligible compared to $\phi_r$ and it does not contribute to the production of curvature perturbation, we will ignore its contribution for the time being. Following \cite{Lemoine:2006sc,Lemoine:2008qj,Gupta:2003jc} we can write the  equations for dimensionless density fluctuations $\Delta_i = \delta \rho_i/\rho_i$ with $i$ denoting radiation $(\gamma)$, baryons $(b)$, anti-baryons $\Bar{b}$, and radial part of the AD field $(\phi_r)$ as follows: 
\begin{eqnarray}
    \frac{d\Delta_{\gamma}}{dN} &=& -\frac{\Gamma_{\phi_r\gamma}}{H} \frac{\Omega_{\phi_r}}{\Omega_{\gamma}}\left(\Delta_{\gamma}  - \Delta_{\phi_r}\right)  - 2 \sum_{i}\Omega_i \Delta_i  - \Psi \left(4 - \frac{\Gamma_{\phi_r \gamma}}{H}\frac{\Omega_{\phi_r}}{\Omega_{\gamma}}\right), \\
    \frac{d\Delta_{b}}{dN} &=& -\frac{\Gamma_{\phi_r b}}{H} \frac{\Omega_{\phi_r}}{\Omega_{b}}\left(\Delta_{b} - \Delta_{\phi_r}\right) - \frac{3}{2} \sum_{i}\Omega_i \Delta_i  - \Psi \left(3 - \frac{\Gamma_{\phi_r b}}{H}\frac{\Omega_{\phi_r}}{\Omega_{b}}\right), \\
    \frac{d\Delta_{\Bar{b}}}{dN} &=& -\frac{\Gamma_{\phi_r \Bar{b}}}{H} \frac{\Omega_{\phi_r}}{\Omega_{\Bar{b}}}\left(\Delta_{\Bar{b}}  - \Delta_{\phi_r}\right) - \frac{3}{2} \sum_{i}\Omega_i \Delta_i  - \Psi \left(3 - \frac{\Gamma_{\phi_r \Bar{b}}}{H}\frac{\Omega_{\phi_r}}{\Omega_{\Bar{b}}}\right), \\
    \frac{d\Delta_{\phi_r}}{dN} &=& -\frac{3}{2} \sum_{i}\Omega_i \Delta_i - \Psi \left(3 + \frac{\Gamma_{\phi}}{H}\right),
\end{eqnarray}
where $\Omega_i = \frac{\rho_i}{\sum_i \rho_i}$ is the dimensionless density of the $i$-th species and $\Psi$ is the Newton potential, and $\Gamma_{\phi} \simeq \Gamma_{\phi_r\gamma}+\Gamma_{\phi_r b}+\Gamma_{\phi_r \Bar{b}}$ is the decay width of the AD/curvaton field, . These equations are applicable after the AD/curvaton field transits to quadratic region of its potential and the baryon number starts to get generated.

The curvature perturbation associated with the $i$-th species is defined as
\begin{equation}
    \zeta_i = - \Psi - H \frac{\delta \rho_i}{\dot{\rho}_i},
\end{equation}
and the baryon and curvaton isocurvature perturbations are defined as
\begin{eqnarray}\label{eq:baryIso}
    \mathcal{S}_B &=& 3 \left(\zeta_b - \zeta_{\gamma}\right),\\ \label{eq:curIso}
    \mathcal{S}_{\phi_r} &=& 3 \left(\zeta_{\phi_r} - \zeta_{\gamma}\right).
\end{eqnarray}
The complex AD field $\Phi$ and its complex conjugate partner $\Phi^{\dagger}$ decay to the baryons and anti-baryons, respectively, with the same decay width $\Gamma_{\phi}$. However, since the baryon asymmetry arises as a result of the difference in the number densities of $\Phi$ and $\Phi^{\dagger}$, the asymmetry is transferred to the thermal plasma. We effectively realize this process by setting $\Gamma_{\phi_r b} \neq \Gamma_{\phi_r \Bar{b}}$. Thus our analysis is approximately the same as the asymmetric decay scenario of \cite{Lemoine:2008qj}.

Following \cite{Lemoine:2008qj} the final curvature perturbation of the radiation can be computed as
\begin{equation}\label{eq:radCurv}
    \zeta_{\gamma}^{(f)} \simeq \left(1 - \Omega_{\phi_r}^{< d}\right) \zeta_{\gamma}^{(i)} + \Omega_{\phi_r}^{<d} \zeta_{\phi_r}^{(i)},
\end{equation}
where $\Omega_{\phi_r}^{<_d}$ is the curvaton fractional energy density just before its decay, and $\zeta_{n}^{(i)}$ is the initial curvature perturbation of the $n$-th species after inflation. It is easy to see that if the curvaton dominates the energy density $(\Omega_{\phi_r}^{<d} \sim 1)$ before its decay, the radiation after curvaton decay inherits the perturbation of the curvaton. Next, to compute the perturbation associated with the baryons, one can define a composite fluid with an energy density $\rho_{comp} = \rho_b - \rho_{\bar{b}} + \Delta B_{b \bar{b}} \rho_{\phi_r}$, with $\Delta B_{b \bar{b}} = \frac{\Gamma_{\phi_r b}-\Gamma_{\phi_r \bar{b}}}{\Gamma_{\phi}}$. Now the perturbation of this  composite fluid can be computed as \cite{Lemoine:2008qj}
\begin{equation}\label{eq:compCurv}
    \zeta_{comp} = \frac{1}{\Omega_b - \Omega_{\bar{b}}+\Delta B_{b \bar{b}} \Omega_{\phi_r}} \left(\Omega_b \zeta_b - \Omega_{\bar{b}} \zeta_{\bar{b}}+ \Delta B_{b \bar{b}}\Omega_{\phi_r} \zeta_{\phi_r}\right),  
\end{equation}
 As this fluid is isolated, $\zeta_{comp}$ is always conserved and after the decay of AD field one can estimate
\begin{equation}\label{eq:baryCurv}
    \zeta_b^{(f)} \simeq \zeta_{comp}^{(f)} \simeq \zeta_{comp}^{(i)}.
\end{equation}
Initially if there is no baryon asymmetry and baryon isocurvature perturbation such that $\Omega_{b}^{(i)} = \Omega_{\bar{b}}^{(i)}$, and  $\zeta_{b}^{(i)} = \zeta_{\bar{b}}^{(i)}$, then again it is easy to see that if the AD/curvaton field dominates the energy density before its decay, the baryons inherits the perturbation of the AD/curvaton field. Now from Eqs. \eqref{eq:compCurv} and \eqref{eq:baryCurv} one can compute $\mathcal{S}_B$ using Eqs.~\eqref{eq:radCurv}, \eqref{eq:baryIso} and \eqref{eq:curIso} as \cite{Lemoine:2008qj}
\begin{equation}
    \mathcal{S}_B = \left(1 - \Omega_{\phi_r}^{<_d}\right) \mathcal{S}_{\phi_r},
\end{equation}
In our scenario where curvaton dominates the energy density before decay (see Table~\ref{Table:nb-s}), the fractional energy density $\Omega_{\phi_r}^{<_d} \sim 1$, which implies $\mathcal{S}_B \sim 0$ due to $\phi_r$, being consistent with the observations.

In the previous analysis even if we have ignored the effect of $\theta$ for the computation of isocurvature perturbation, it can be responsible for uncorrelated isocurvature perturbation. The contribution of the angular part $\theta$ to the baryon isocurvature perturbation can be related to the temperature fluctuations as \cite{Hu:1994uz,Barrie:2021mwi}
\begin{eqnarray}\nonumber
	\left(\frac{\delta T}{T}\right)_{\theta} &=& -\frac{2}{5} \frac{\Omega_B}{\Omega_m} \frac{\delta n_B}{n_B} \\ \label{eq:isoTemp}
	&=& -\frac{4}{5} \frac{\Omega_B}{\Omega_m} \cot(2 \theta_I) \delta \theta.
\end{eqnarray}
The fluctuations of the angular direction $\delta \theta$ can be computed as $\delta \theta = \frac{H_I}{2 \pi} \frac{1}{\phi_{r,I}}$. The adiabatic temperature fluctuations can be computed as
\begin{equation}
	\left \langle \left(\frac{\delta T}{T}\right)_{adi}^2 \right \rangle = \frac{1}{20} P_{\zeta}.
\end{equation}
With $\frac{\Omega_B}{\Omega_m} = 0.16$ we can compute the ratio between the angular isocurvature to adiabatic perturbations as
\begin{equation}
	\alpha_{iso} = \frac{\left \langle \left(\frac{\delta T}{T}\right)_{\theta}^2 \right \rangle}{\left \langle \left(\frac{\delta T}{T}\right)_{adi}^2 \right \rangle} = \frac{320}{25 P_{\zeta}} (0.16)^2 \cot^2(2 \theta_I) \frac{H_I^2}{4 \pi^2} \frac{1}{\phi_{r,I}^2}.
\end{equation}
The current bound on the ratio is $\alpha_{iso} \leq 1.9 \times 10^{-2}$ \cite{Planck:2018jri}. With the benchmark values in Table \ref{Table:nb-s} we get the following constraints on the initial value of $\theta_I$:
\begin{eqnarray*}
	\theta_I \gtrsim 0.74 \text{ for } H_I = 10^{-5}, \phi_{r,I} = 1.17 \times 10^{-2} ,\\
	\theta_I \gtrsim 0.72  \text{ for } H_I = 3.21 \times 10^{-6}, \phi_{r,I} = 5.54 \times 10^{-3} .	
\end{eqnarray*}
Ofcourse the initial value of the angular direction can not be arbitrary, but the above bounds are consistent with our choice of $Q_{\phi}\sin(2 \theta_I) \sim 1$. This analysis suggests that isocurvature perturbation will be observationally consistent for sufficient domination of the radial part before decay and a suitable initial choice of the angular part.
\bibliographystyle{JHEP}
 \bibliography{biblio.bib}

\end{document}